\documentclass[showpacs, showkeys, amsmath,amssymb]{revtex4}
\usepackage[utf8]{inputenc}
\usepackage{ulem}
\usepackage{amsmath}
\usepackage{amssymb}
\usepackage{amsfonts}
\usepackage{array}
\usepackage{graphicx}
\usepackage{mathrsfs}
\usepackage{multirow}
\usepackage{xcolor}
\usepackage{slashed}
\usepackage{extarrows}
\usepackage{bbm}
\usepackage{indentfirst}

\newcommand{\ie}{{\it i.e.}}

\newcommand{\etl}{{\it et.al.}}

\newcommand{\bea}{\begin{eqnarray*}}
\newcommand{\eea}{\end{eqnarray*}}
\newcommand{\beao}{\begin{eqnarray}}
\newcommand{\eeao}{\end{eqnarray}}

\def\cmag{\color{magenta}}

\providecommand{\Journal}[4] {#1 {\bf#2}, #4 (#3)}
\providecommand{\PRL}{Phys. Rev. Lett.} %
\providecommand{\PRD}{Phys. Rev. D} %
\providecommand{\JCAP}{JCAP} %
\providecommand{\PRSLA}{Proc. R. Soc. Lond. A.}
\providecommand{\RMP}{Rev. Mod. Phys.}

\begin{document}
\title{Analytical solutions of CPT-odd Maxwell equations in Schwarzschild spacetime}
		
\author{Hao Wang${}^{a, b}$}
\author{Zhi Xiao${}^{a, b}$}
\email{spacecraft@pku.edu.cn}
\author{Bing Sun${}^c$}
\affiliation{${}^{a}$Department of Mathematics and Physics, North China Electric Power University, Beijing 102206, China}
\affiliation{${}^{b}$Hebei Key Laboratory of Physics and Energy Technology, North China Electric Power University, Baoding 071000, China}
\affiliation{${}^{c}$Department of Basic Courses, Beijing University of Agriculture,\\ \small Beijing 102206, China}
	
\begin{abstract}
In this work, we present the CPT-violating (CPTV) Maxwell equations in curved spacetime using the Newman-Penrose (NP) formalism. We obtain a semi-analytical solution to the Maxwell equations in Schwarzschild spacetime under the assumption that the CPT-odd $\left(k_{AF}\right)^\mu$ term exhibits spherical symmetry in the Schwarzschild background. Retaining only terms up to linear order in the $\left(k_{AF}\right)^\mu$ coefficient, we obtain perturbative solutions by treating the solutions of the Lorentz-invariant Maxwell equations as the zeroth-order approximation and incorporating the $\left(k_{AF}\right)^\mu$ terms as an additional source term alongside the external charge current. Each resulting NP scalar field can be factorized into two components: the radial component is expressed in terms of hypergeometric functions, while the angular component is described by spin-weighted spherical harmonics.
\end{abstract}
	
\keywords{Maxwell equations, CPT violation, Null formalism}
	
\maketitle

\section{Introduction}
Lorentz symmetry (LS) is a fundamental symmetry in both general relativity (GR) \cite{SWeinGrav} and the standard model \cite{SWeinQFT1} of particle physics.
Some candidate theories of quantum gravity, such as certain formulations of loop quantum gravity  \cite{loop1,loop2}, and modifications of string theory \cite{string,SMEa},
allow for small deviations from exact Lorentz invariance at very high energies, which could, in some scenarios, lead to tiny Lorentz-violating effects at lower energies.
However, Lorentz-violating signals at energy scales accessible in high-energy astrophysical observations ( $\sim 10^{11} \mathrm{GeV}$ \cite{UHE of CR}) are expected to be extremely small
and are generally suppressed by a small ratio involving the Planck scale $1.22 \times 10^{19} \mathrm{GeV}$, as suggested by dimensional analysis and observational constraints.

The tiny Lorentz violation (LV) effects may accumulate over long distances and at high energies in certain models, making them potentially detectable through terrestrial observations.
Additionally, some experiments and astrophysical observations can test processes that are strictly forbidden in standard Lorentz-invariant (LI) physics but may occur in LV scenarios,
such as vacuum birefringence \cite{vacuum-birefrin1,vacuum-birefrin2}, photon decay  \cite{photon-decay}, and photon splitting \cite{photon-splitting}.
These observations have placed stringent constraints on LV with high precision, particularly through high-energy observatories such as Pierre Auger and LHAASO \cite{Auger18,LHAASO24}.
Moreover, most of these observatories primarily rely on multi-wavelength observations and long-distance photon propagation to study astrophysical events.

As a comprehensive framework of effective field theory,
the Standard Model Extension is capable of describing both small deviations from LS in flat spacetime \cite{SMEa,SMEb}
and violations of local LS in gravitational contexts  \cite{SMEg},
thus encompassing both high-energy phenomena in flat spacetime \cite{SMEa} and gravitational effects or particle motions in curved spacetime  \cite{SME-gra,SME-mat}.
In recent years, there has been increasing interest in probing LV in astrophysics through observations of CMB photons \cite{CMBpol-LV1, CMBpol-LV2}, neutrinos \cite{neut-LV},
and gravitational waves \cite{GW-LV}.
A natural question is whether interesting LV effects manifest in intrinsically curved spacetime.
For the constraints on LV from studying the cosmological propagation of GRB or CMB photons,
a key assumption is that spacetime is described by the Friedmann-Lemaître-Robertson-Walker (FLRW) metric,
which is curved but conformally flat.
Here, we investigate LV electrodynamics in the simplest non-conformally flat curved spacetime: the Schwarzschild metric.

There has been a growing number of studies on photon behavior in curved spacetime,
particularly following the successful observation of black hole (BH) images by the Event Horizon Telescope (EHT) Collaboration \cite{EHTBHI}.
The photon sphere and the subrings at the edge of the black hole shadow may encode crucial information about potential new physics beyond GR \cite{UISBHPR}.
This highlights the necessity of studying photon behavior in curved spacetime near BHs.

As a first attempt, we aim to investigate the asymptotic behavior of CPT-violating (CPTV) photons in the Schwarzschild geometry using the null formalism.
Intuitively, employing null formalism to study massless particles is, in some sense,
analogous to describing the motion of massive particles-such as a gyroscope-using an orthonormal tetrad in its instantaneous rest frame,
despite the fact that massless particles do not possess a rest frame.
However, the underlying principle remains the same: null tetrads naturally accommodate massless particles that follow null trajectories.
Moreover, the null formalism offers unique advantages in analyzing the asymptotic behavior of massless particles, particularly photons in this context.
It significantly simplifies the description of both the tangent of the null geodesic for photons and gravitons, as well as their polarization states \cite{Dicong2017}.
Consequently, it provides a coordinate-independent framework for studying photon dynamics with a clear geometric interpretation,
especially in highly curved spacetimes.
Notably, it also offers a physically transparent decomposition of the Faraday tensor into ingoing, outgoing, and Coulomb modes.

In the presence of CPT or Lorentz violation, contrary to conventional expectations,
the asymptotic behavior of photons may be qualitatively altered  \cite{CPTVAsymp}.
For instance, vacuum birefringence can be understood in terms of the modified topology of the light cone structure,
where different helicities of CPT-odd photons experience distinct causal cones \cite{Itin2007}.

Pioneering works on exact solutions for LI photon fields as perturbations in given background geometries
include vacuum solutions for photon fields in the Kerr spacetime \cite{Chand76},
as well as solutions for a point charge near a Schwarzschild BH \cite{Wald-Schw} and a Kerr BH \cite{Wald-Kerr}, among others.
Bičák et al. have studied photon fields in curved spacetime within the Newman-Penrose (NP) framework \cite{NP1962}, including the Schwarzschild \cite{LI-electro-NP},
ReissnerNordström (R-N) \cite{RNNP}, and Kerr \cite{kerrNP} black hole backgrounds.
As a preliminary attempt, we study the behavior of CPT-odd photons in the Schwarzschild geometry following a similar approach.

In Sec. 2, we review the Newman-Penrose (NP) formalism and discuss some earlier studies on the LI Maxwell equations in curved spacetime using the NP framework.
Then, we examine the CPT-violating Maxwell equations within the NP formalism in curved spacetime.
Next, we present a method to solve the coupled Maxwell equations and provide some special solutions in Sec. 3.
In the last section, we summarize our results and provide a short conclusion.
In this work, the signature of the metric tensor $g_{\mu \nu}$ is chosen to be $(+,-,-,-)$,
and we use geometric units with $\epsilon_0=\mu_0=c=G=\hbar=1$.
The notation conventions are as follows: spacetime indices are represented by Greek letters such as $\mu, \nu, \rho$,
while null tetrad indices are represented by Latin letters such as $a, b, c$.

\section{field equations and solutions}
We study LV (more specifically CPTV) photon behavior within the photon sector of the minimal SME \cite{SMEa}.
The action is given by
\begin{equation}
S
= \int d^4x\sqrt{-g}
	\left[-\frac{1}{4}F_{\mu \nu}F^{\mu \nu} + (k_{AF})_\alpha A_\beta
		\tilde{F}^{\alpha \beta}-J^{\mu}A_{\mu}\right],
\end{equation}
where $\tilde{F}^{\alpha \beta}=\frac{1}{2}\epsilon ^{\mu \nu \alpha \beta } F_{\mu \nu }$, and $(k_{AF})_\alpha$ is the CPTV coefficient \cite{SMEa} \cite{CFJ1990}. The coefficients $(k_{AF})_\alpha$ are real and have mass dimension one. The equation of motion follows as
\begin{equation}
    \nabla _\mu F^{\mu \nu}+2(k_{AF})_{\mu } \widetilde{F}^{\mu \nu} =J^{\nu}.\label{EOM-CPT-odd}
\end{equation}
Using the null tetrad $e_a^{\ \mu} =\left( l^{\mu},n^{\mu},m^{\mu},\bar{m}^{\mu}\right), a=1,2,3,4,$  correspond respectively to $l,n,m, \bar{m}$, the electromagnetic field tensor $F_{\mu\nu}$ can be decomposed into three complex Newman-Penrose (NP) scalars,
\begin{equation}
{\begin{aligned}&
\Phi_{0}=F_{13}=F_{\mu \nu}e_{1}{}^{\mu}e_{3}{}^{\nu}=F_{\mu \nu} l^{\mu} m^{\nu}, \\&
\Phi_{1}=\frac{1}{2}\left(F_{12}+F_{43}\right)=\frac{1}{2} F_{\mu \nu}\left(e_{1}{}^{\mu}e_{2}{}^{\nu}+e_{4}{}^{\mu}e_{3}{}^{\nu} \right)=\frac{1}{2} F_{\mu \nu}\left(l^{\mu} n^{\nu}+\bar{m}^{\mu} m^{\nu}\right) ,\\&
\Phi_{2}=F_{42}=F_{\mu \nu}e_{4}{}^{\mu}e_{2}{}^{\nu}=F_{\mu \nu} \bar{m}^{\mu} n^{\nu}.
\end{aligned}}
\end{equation}
Conversely, the electromagnetic field tensor $F^{\mu\nu}$ can be expressed as
\begin{equation}
F_{\mu \nu}=2\left\{\Phi_1\left(n_{[\mu} l_{\nu]}+m_{[\mu} \bar{m}_{\nu]}\right)+\Phi_2 l_{[\mu} m_{\nu]}+\Phi_0 \bar{m}_{[\mu} n_{\nu]}\right\}+\text{ c.c. },
\end{equation}
where $a_{[\mu}b_{\nu]}:=\frac{1}{2}(a_{\mu}b_{\nu}-a_{\nu}b_{\mu})$, and $``\text{c.c.}"$ denotes the complex conjugate.

This work primarily focuses on photon fields propagating in the vicinity of a Schwarzschild black hole (BH),
with the line element
\begin{equation}
ds^{2}=g(r) dt^{2}-g^{-1}(r) d r^{2}-r^{2}\left(d \theta^{2}+\sin ^{2} \theta d \varphi^{2}\right),\label{Sch-gauge}
\end{equation}
where $g(r)=1-\frac{2 M}{r}$. The corresponding null tetrad basis vectors are given by
\begin{equation}
\begin{aligned}\label{nullV}
l^{\mu}&=\left(g(r)^{-1}, 1,0,0\right), &n^{\mu}=\frac{1}{2}(1,-g(r), 0,0),~m^{\mu}&=\frac{1}{\sqrt{2} r}(0,0,1, i \csc \theta), ~\bar{m}^{\mu}=\frac{1}{\sqrt{2} r}(0,0,1,-i \csc \theta)
\end{aligned}
\end{equation}
with their corresponding covariant components given by
\begin{equation}
{\begin{aligned}
l_{\mu}&=(1,-g(r)^{-1},0,0), &n_{\mu}=\frac{1}{2}\left(g(r), 1, 0,0\right),~ m_{\mu}&=\frac{r}{\sqrt{2}}(0,0,-1,-i \sin \theta),~ \bar{m}_{\mu}=\frac{r}{\sqrt{2}}(0,0,-1, i \sin \theta),
\end{aligned}}
\end{equation}
In terms of the null contravariant vectors, we project the derivatives into null directions:
\begin{equation}
\begin{aligned}&
D=l^{\mu} \nabla_{\mu}=g(r)^{-1} \partial _t+\partial _r, & \delta=m^{\mu} \nabla_{\mu}=\frac{1}{\sqrt{2}r}\left(\partial_{\theta}+i \csc \theta \partial_ \varphi\right), \\ &
\Delta=n^{\mu} \nabla_{\mu}=\frac{1}{2} \partial _t-\frac{1}{2} g(r) \partial _r, & \bar{\delta}=\bar{m}^{\mu} \nabla_{\mu}=\frac{1}{\sqrt{2}r}\left(\partial_{\theta}-i \csc \theta \partial_ \varphi\right),
\end{aligned}\label{covariant-D-Os}
\end{equation}
following Ref. \cite{NP1962}.
To simplify the calculations, we assume that the CPTV coefficient is spherically symmetric, consistent with the Schwarzschild background, and restrict our analysis to stationary electromagnetic fields. Consequently, the directional derivatives reduce to $D=\partial_r$ and $\Delta=-\frac{1}{2} g(r) \partial _r$.
In other words, we focus on the behavior of static electric and magnetic fields in the presence of the CPT-odd term.

Similarly, we define the spin coefficients as $\gamma_{abc} \equiv e_a^{\ \mu} e_{b\mu;\nu} e_c^{\ \nu}$, following the conventions in Ref. \cite{Chandrasekhar}.
In the Schwarzschild metric, the nonzero spin coefficients are given by
\begin{equation}
\begin{aligned}
\rho&\equiv\gamma_{314}=-\frac{1}{r},\quad \mu\equiv\gamma_{243}=-\frac{1}{2 r}\left(1-\frac{2 M}{r}\right),\quad \gamma\equiv\frac{1}{2}\left(\gamma_{212}+\gamma_{342}\right)=\frac{M}{2 r^{2}}, \\ &
\alpha\equiv\frac{1}{2}\left(\gamma_{214}+\gamma_{344}\right)=-\frac{1}{2 \sqrt{2} r} \cot \theta, \quad
\beta\equiv\frac{1}{2}\left(\gamma_{213}+\gamma_{343}\right)=\frac{1}{2 \sqrt{2} r} \cot \theta.
\end{aligned}\label{spin-coefficinets}
\end{equation}
It is important to note that the Newman–Penrose (NP) formalism employed here may not be fully applicable
within a generic Lorentz-violating theory.
In this work, we have implicitly adopted the test particle assumption,
wherein the background metric is assumed to remain unaffected by the Lorentz-violating matter fields ---
specifically, the electromagnetic fields under consideration.
Within this framework, the use of a complete and quasi-orthonormal null tetrad remains appropriate,
as it effectively captures the essential features of the quasi-null wavefront associated with CPT-violating electromagnetic fields.
This is because the null tetrad can be regarded as a natural choice for describing massless particles.

However, for a more rigorous treatment --- particularly when the back-reaction of matter fields
on the spacetime metric is taken into account --- the standard null tetrad may no longer suffice.
In such cases, it may be necessary to generalize the framework, for example, by employing a quasi-null tetrad,
as used in the analysis of gravitational wave polarization \cite{CW-PolarizationGW2018}.

\subsection{CPTV Maxwell Equations in the NP Formalism}
The LI Maxwell equations in the NP formalism have been derived in the appendix of Ref. \cite{NP1962} and in Chapter 1.8 of Chandrasekhar’s textbook \cite{Chandrasekhar}.
For the CPTV contribution, the term $(k_{AF})_{\mu} \widetilde{F}^{\mu\nu}$ in Eq. (\ref{EOM-CPT-odd}) can be projected onto the null tetrad basis as:
$(k_{AF})_{a } \widetilde{F}^{a b} $, where $\widetilde{F}^{a b}\equiv\frac{1}{2}\epsilon _{a b c d } F^{c d}$ and $ (k_{AF})_{a} \equiv(k_{AF})_{\mu}\mathbf{e}_a ^{\ \mu} \ (a=1, 2, 3, 4$), which are tetrad components of the CPTV coefficient $(k_{AF})_{\mu}$, and for simplicity, we define $(k_{AF})^{a}\equiv  k^a $.
As an example with $b=1$, we obtain:
\begin{equation}
\begin{aligned}
2(k_{AF})^{a } \widetilde{F}_{a 1} = (k_{AF})^{a} \epsilon_{a  1 c d } F^{c d}= 2i \left[ -k^2 (\Phi_1-\bar{\Phi}_1) -k^3 \Phi_0 +k^4 \bar{\Phi}_0 \right].
\end{aligned}
\end{equation}
Here, we use $\epsilon_{1234} =i$, which follows from the definition of the complex null tetrad given in (\ref{nullV}).
The CPTV Maxwell equations in NP form are then given by:
\begin{equation}
\begin{aligned}
&(D-2 \rho) \Phi_1-(\bar{\delta}+\pi-2 \alpha) \Phi_0+\kappa \Phi_2=\frac{1}{2} J_l+i\left[ k^{4} \bar{\Phi}_{0}- k^{3} \Phi_{0}- k^{2}\left(\Phi_{1}-\bar{\Phi}_{1}\right)\right], \\&
(\delta-2 \tau) \Phi_1-(\Delta+\mu-2 \gamma) \Phi_0+\sigma \Phi_2 =\frac{1}{2} J_m+i\left[k^{4}\left(\Phi_{1}+\bar{\Phi}_{1}\right)+k^{2} \bar{\Phi}_{2}+ k^1 \Phi_{0}\right], \\&
(D-\rho+2 \varepsilon) \Phi_2-(\bar{\delta}+2 \pi) \Phi_1+\lambda \Phi_0=\frac{1}{2} J_{\bar{m}} -i\left[k^{3}\left(\Phi_{1}+\bar{\Phi}_{1}\right)+ k^{2} \Phi_{2}+k^{1} \bar{\Phi}_{0}\right], \\&
(\delta-\tau+2 \beta) \Phi_2-(\Delta+2 \mu) \Phi_1+\nu \Phi_0 =\frac{1}{2} J_n+ i\left[k^{4} \Phi_{2}-k^{3} \bar{\Phi}_{2}+k^{1}\left(\Phi_{1}-\bar{\Phi}_{1}\right)\right].
\end{aligned}
\label{NP_LV_Maxwell_eq}
\end{equation}
where $J_l = l_\mu j^\mu$, $J_n = n_{\mu} j^\mu$, {\it etc}.

Since we assume that the CPTV coefficient $k^\mu$ is spherically symmetric, a simple example is to consider only $k^t \neq 0$.
Given that $k^t = k^{1} l^{t} + k^{2} n^{t} \neq 0$, this implies that $k^{1}$ and $k^{2}$ cannot be zero, while $k^3 = k^4 = 0$.
By substituting the spin coefficients from Eq. \eqref{spin-coefficinets} and the differential operators from Eq. (\ref{covariant-D-Os}), Eq. (\ref{NP_LV_Maxwell_eq}) simplifies to:
\begin{equation}
\begin{aligned}&
{\left(\frac{\partial}{\partial r}+\frac{2}{r}\right) \Phi_{1}+\frac{1}{\sqrt{2} r} \bar{\eth} \Phi_{0}=\frac{1}{2} J_l-i k^2\left(\Phi_{1}-\bar{\Phi}_{1}\right)}, \\&
-\frac{1}{\sqrt{2} r} \bar{\eth} \Phi_{1}+\frac{1}{2}\left[\left(1-\frac{2 M}{r}\right) \frac{\partial}{\partial r}+\frac{1}{r}\right] \Phi_{0}=\frac{1}{2} J_m+i k^2\bar{\Phi}_{2}+i k^1 \Phi_{0},\\&
(\frac{\partial}{\partial r}+\frac{1}{r}) \Phi_{2}+\frac{1}{\sqrt{2} r} \bar{\eth} \Phi_{1}= \frac{1}{2} J_{\bar{m}} -i k^2 \Phi_{2}-i k^1 \bar{\Phi}_{0},\\&
-\frac{1}{\sqrt{2} r} \eth \Phi_{2}+\frac{1}{2}\left(1-\frac{2 M}{r}\right)\left(\frac{\partial}{\partial r}+\frac{2}{r}\right) \Phi_{1}=\frac{1}{2} J_n+i k^1\left(\Phi_{1}-\bar{\Phi}_{1}\right),
\end{aligned} \label{spin-s LV Maxwell}
\end{equation}
where the differential operators $\eth$ and $\bar{\eth}$ are defined as:
\begin{eqnarray}&&
\eth \eta \equiv -(\sin \theta)^s\left[\frac{\partial}{\partial \theta}+\frac{\mathrm{i}}{\sin \theta} \frac{\partial}{\partial \varphi}\right](\sin \theta)^{-s} \eta, \\&&
\bar{\eth} \eta \equiv -(\sin \theta)^{-s}\left[\frac{\partial}{\partial \theta}-\frac{\mathrm{i}}{\sin \theta} \frac{\partial}{\partial \varphi}\right](\sin \theta)^s \eta.
\end{eqnarray}
The eigenfunctions of $\eth$ and $\bar{\eth}$ are the spin-weighted spherical harmonics $_{s}Y_{l m}$.
Here, $s=1, 0, -1$ correspond to the spin weights of $\Phi_0, \Phi_1, \Phi_2$, respectively.
For $s=0$, $_{0}Y_{l m} = Y_{l m}$ are the standard spherical harmonics, and the indices $s, l, m$ satisfy $|m|\leq l$ and $|s| \leq l$.
For further details on the definition and properties of spin-weighted spherical harmonics, see Appendix A or Ref. \cite{spin-s}.

\section{Solutions for a Given Source}	
To solve Eqs.~(\ref{spin-s LV Maxwell}), we adopt the Teukolsky approach \cite{Teukolsky-method}.
The key is to utilize the commutation relations \cite{NP1962} between differential operators (\ref{covariant-D-Os}) and spin coefficients (\ref{spin-coefficinets}) to decouple the four coupled Maxwell equations.
Acting with a combination of the operators on Eq.~(\ref{spin-s LV Maxwell}),
we obtain a set of partially decoupled equations:
\begin{subequations}
\begin{align}
&(D-3 \rho)(\Delta+\mu-2 \gamma) \Phi_{0}-\delta(\bar{\delta}-2 \alpha) \Phi_{0}=\frac{1}{2} J_{0}  -i\left[ k^2  \delta\left(\Phi_{1}-\bar{\Phi}_{1}\right)+(D-3 \rho)  k^2 \bar{\Phi}_{2}+(D-3 \rho) k^1 \Phi_{0}\right], \\ \nonumber\\
&(D-2 \rho)(\Delta+2 \mu) \Phi_{1}-(\delta+2 \beta) \bar{\delta} \Phi_{1}=\frac{1}{2} J_{1}  -i\left[(\delta+2 \beta)  k^2 \Phi_{2}+(\delta+2 \beta) k^1 \bar{\Phi}_{0}+(D-2 \rho) k^1\left(\Phi_{1}-\bar{\Phi}_{1}\right)\right], \\\nonumber\\
&(\Delta+3 \mu)(D-\rho) \Phi_{2}-\bar{\delta}(\delta+2 \beta) \Phi_{2}=\frac{1}{2} J_{2} -i\left[ k^2(\Delta+3 \mu) \Phi_{2}+k^1(\Delta+3 \mu) \bar{\Phi}_{0}+\bar{\delta} k^1\left(\Phi_{1}-\bar{\Phi}_1\right)\right].
\end{align}
\label{TParTeuDec}
\end{subequations}
where
\begin{equation}
\left\{ \begin{aligned}
& J_0:=\delta J_l-(D-3 \rho) J_m, \\
& J_1:=(\delta+2\beta) J_{\bar{m}}-(D-2\rho) J_n, \\
& J_2:=(\Delta+3 \mu) J_{\bar{m}}-\bar{\delta}J_n.
\end{aligned} \right.
\end{equation}
For later convenience, we define:
$\sum_{lm} \equiv \sum_{l=1}^{\infty} \sum_{m=-l}^{l}, $
Since spherical symmetry is preserved in at least a special preferred reference frame, we expand the three complex scalars using spin-weighted spherical harmonics:
\begin{equation}
\left\{\begin{aligned}
\Phi_{0} &= \sum_{l m} R_{0|l m}(r)\  _1Y_{l m}(\theta, \varphi), \\
\Phi_{1} &= \sum_{l m} R_{1|l m}(r)\  _0Y_{l m}(\theta, \varphi) + R_{1|0 0}(r)\  _0Y_{00}(\theta, \varphi), \\
\Phi_{2} &= \sum_{l m} R_{2|l m}(r)\ _{-1}Y_{l m}(\theta, \varphi).
\end{aligned}\right.
\label{formal solution}
\end{equation}
Inspection of Eqs.~(\ref{TParTeuDec}) reveals that in the absence of CPTV coefficients, the three equations are decoupled. Since the CPTV coefficients are experimentally constrained to be very tiny,
say $|k_{AF}|\le10^{-44}$GeV, \cite{CMBLV-N, Cherenkov, Opticalpolarization}, we may treat the CPTV terms on the right-hand side of Eqs.~(\ref{TParTeuDec}) as perturbations.

Thus, the radial functions can be expanded in powers of CPTV coefficients $k^1$ and $k^2$ as
\begin{equation}
R_{a|l m}(r) = R_{a|l m}^{(0)} + R_{a|l m}^{(1)} + R_{a|l m}^{(2)} + \cdots, \quad a=0,1,2,
\end{equation}
where the superscripts ``(0)", ``(1)", and so on indicate the corresponding order of $k^1$ and $k^2$.
For example, $R_{a|l m}^{(0)}$ corresponding to zero-th order function without LV correction.
The expansions of the NP scalars are thus given by:
\begin{equation}
\left\{\begin{aligned}
\Phi_{0} &= \Phi_{0}^{(0)}+\Phi_{0}^{(1)}+\cdots= \sum_{lm} \left(R_{0|l m}^{(0)}+R_{0|l m}^{(1)}+\cdots\right){ }_{1}Y_{l m}, \\
\Phi_{1} &= \Phi_{1}^{(0)}+\Phi_{1}^{(1)}+\cdots=\sum_{lm}{}^{\prime}\left(R_{1|lm}^{(0)}+R^{(1)}_{1|lm}+\cdots\right)  { }_{0}Y_{l m}, \\
\Phi_{2} &= \Phi_{2}^{(0)}+\Phi_{2}^{(1)}+\cdots= \sum_{lm}\left(R_{2|l m}^{(0)}+R_{2|l m}^{(1)}+\cdots\right){ }_{-1} Y_{l m}.
\end{aligned} \right.
\label{01formal solution}
\end{equation}
Keeping only the linear-order terms in the CPTV coefficients, Eqs.~(\ref{TParTeuDec}) can be separated into two sets: the zeroth-order Lorentz-invariant (LI) equations,
\begin{equation}
\begin{aligned}
(D-3 \rho)(\Delta+\mu-2 \gamma) \Phi_{0}^{(0)}-\delta(\bar{\delta}-2 \alpha) \Phi_{0}^{(0)} &=\frac{1}{2} J_{0}, \\
(D-2 \rho)(\Delta+2 \mu) \Phi_{1}^{(0)}-(\delta+2 \beta) \bar{\delta} \Phi_{1}^{(0)} &=\frac{1}{2} J_{1}, \\
(\Delta+3 \mu)(D-\rho) \Phi_{2}^{(0)}-\bar{\delta}(\delta+2 \beta) \Phi_{2}^{(0)} &=\frac{1}{2} J_{2},
\end{aligned}
\label{NPLV-Maxw-0th}
\end{equation}
and the first-order equations with linear CPTV corrections,
\begin{equation}
\begin{aligned}
& (D-3 \rho)(\Delta+\mu-2 \gamma) \Phi_0^{(1)}-\delta(\bar{\delta}-2 \alpha) \Phi_0^{(1)}
=-i\left[ k^2 \delta\left(\Phi_1^{(0)}-\bar{\Phi}_1^{(0)}\right)+(D-3 \rho)  k^2 \bar{\Phi}_2^{(0)}+(D-3 \rho) k^1 \Phi_0^{(0)}\right],  \\
& (D-2 \rho)(\Delta+2 \mu) \Phi_1^{(1)}-(\delta+2 \beta) \bar{\delta} \Phi_1^{(1)}
=-i\left[(\delta+2 \beta)  k^2 \Phi_2^{(0)}+(\delta+2 \beta) k^1 \bar{\Phi}_0^{(0)}+(D-2 \rho) k^1\left(\Phi_1^{(0)}-\bar{\Phi}_1^{(0)}\right)\right],  \\
& (\Delta+3 \mu)(D-\rho) \Phi_2^{(1)}-\bar{\delta}(\delta+2 \beta) \Phi_2^{(1)}
=-i\left[ k^2(\Delta+3 \mu) \Phi_2^{(0)}+k^1(\Delta+3 \mu) \bar{\Phi}_0^{(0)}+\bar{\delta} k^1\left(\Phi_1^{(0)}-\bar{\Phi}_1^{(0)}\right)\right].
\end{aligned}\label{NPLV-Maxw-1st}
\end{equation}
The zeroth-order equations indicate that the external charge and current serve as sources for the zeroth-order NP complex scalars, which are linear combinations of the components of the Faraday tensor. Similarly, the first-order equations show that the zeroth-order Faraday fields act as sources for the first-order CPTV corrections in the Faraday tensor.

We begin by solving the zeroth-order decoupled equations given in Eq.~(\ref{NPLV-Maxw-0th}). By utilizing the orthogonality relations of spin-weighted spherical harmonics (\ref{spheri-harmon-identy}) and substituting Eqs.~(\ref{covariant-D-Os}), (\ref{spin-coefficinets}) and (\ref{formal solution}) into Eq.~(\ref{NPLV-Maxw-0th}), we obtain the following radial equations:
\begin{equation}
\begin{aligned}
& r(r-2M) R_{0|l m}^{(0)}{}'' + 4(r-M) R_{0|l m}^{(0)}{}' - (l-1)(l+2) R_{0|l m}^{(0)} = - J_{0|l m}, \\
& r(r-2M) R_{1|l m}^{(0)}{}'' + 4(r-\frac{3}{2}M) R_{1|l m}^{(0)}{}' - (l-1)(l+2) R_{1|l m}^{(0)} = - J_{1|l m}, \\
& r(r-2M) R_{2|l m}^{(0)}{}'' + 4(r-2M) R_{2|l m}^{(0)}{}' - \left[(l-1)(l+2) - \frac{4M}{r} \right] R_{2|l m}^{(0)} = - J_{2|l m}.
\end{aligned}
\label{LVeq_of_radial_part0}
\end{equation}
Here, $R(r)''$ and $R(r)'$ denote the second- and first-order derivatives with respect to $r$, respectively. The source terms are given by:
\begin{equation}
\left\{\begin{aligned}
J_{0|l m} &= \int J_{0}(r, \theta, \varphi)\ { }_{1}\bar{Y}_{l m}(\theta, \varphi)\, r^{2} d \Omega, \\
J_{1|l m} &= \int J_{1}(r, \theta, \varphi)\ { }_{0}\bar{Y}_{l m}(\theta, \varphi)\,  r^{2} d \Omega, \\
J_{2|l m} &= \int J_{2}(r, \theta, \varphi)\ { }_{-1}\bar{Y}_{l m}(\theta, \varphi)\,  r^{2} d \Omega.
\end{aligned}\right.
\end{equation}
In the absence of source terms in Eq.~(\ref{LVeq_of_radial_part0}), introducing the variable transformation $x \equiv r/(2M)$, the homogeneous equations can be rewritten in the standard form of hypergeometric equations:
\begin{subequations}
\begin{align}
& x(x-1) R_{0|l m}^{(0)}{}'' + (4x-2) R_{0|l m}^{(0)}{}' - (l-1)(l+2) R_{0|l m}^{(0)} = 0, \label{HGeq0} \\
& x(x-1) R_{1|l m}^{(0)}{}'' + (4x-3) R_{1|l m}^{(0)}{}' - (l-1)(l+2) R_{1|l m}^{(0)} = 0, \label{HGeq1} \\
& x(x-1) R_{2|l m}^{(0)}{}'' + (4x-4) R_{2|l m}^{(0)}{}' - \left[(l-1)(l+2) - \frac{2}{x} \right] R_{2|l m}^{(0)} = 0. \label{HGeq2}
\end{align}
\end{subequations}

The general solutions of the hypergeometric equations are:
\begin{equation}
\begin{aligned}
R^{(0)}_{0|l m} & = a^{(0)}_{l m} R_{0|l}^{(\mathrm{I})} + b^{(0)}_{l m} R_{0|l}^{(\mathrm{II})}, \\
R^{(0)}_{1|l m} & = c^{(0)}_{l m} R_{1|l}^{(\mathrm{I})} + d^{(0)}_{l m} R_{1|l}^{(\mathrm{II})}, \\
R^{(0)}_{2|l m} & = e^{(0)}_{l m} R_{2|l}^{(\mathrm{I})} + f^{(0)}_{l m} R_{2|l}^{(\mathrm{II})}.
\end{aligned}
\label{GS_of_0R}
\end{equation}
For $l \neq 0$, the linearly independent solutions for $R_{a|l m}$ $(a=0,1,2)$ are:
\begin{eqnarray}
R_{0|l}^{(\mathrm{I})}  &=& F(1-l, l+2,2 ; x), \quad
R_{0|l}^{(\mathrm{II})} = (-x)^{-l-2} F\left(l+1, l+2,2l+2 ; x^{-1}\right), \label{solution_of_radial_partb} \\
R_{1|l}^{(\mathrm{I})}  &=& F(1-l, l+2,3 ; x), \quad
R_{1|l}^{(\mathrm{II})} = (-x)^{-l-2} F\left(l, l+2,2l+2 ; x^{-1}\right), \label{solution_of_radial_partd} \\
R_{2|l}^{(\mathrm{I})}  &=& x^{-1} F(-l, l+1,2 ; x), \quad
R_{2|l}^{(\mathrm{II})} = (-x)^{-l-2} F\left(l+1, l,2l+2 ; x^{-1}\right). \label{solution_of_radial_partf}
\end{eqnarray}
For $l=0$, since $R_{0|l}$ and $R_{2|l}$ correspond to spin-weight $s=\pm1$ and the spin-weighted functions
are only defined for $l\ge|s|$, $R_{a|0}$ is not defined except for $a=1$, which corresponds to spin-weight $s=0$.
The linearly independent solutions of $R_{1|0}$ are:
\begin{equation}
R_{1|0}^{(\mathrm{I})} = x^{-2}\ln(x-1) + x^{-1}, \quad R_{1|0}^{(\mathrm{II})} = x^{-2}.
\end{equation}
To fully characterize the solutions, we examine their asymptotic behaviors at both spatial infinity and near the event horizon. The asymptotic expressions of the solutions in the far-field regime are:
\begin{equation}
R^{(\mathrm{I})}_{a|l} \sim x^{-1+l}, \quad R^{(\mathrm{II})}_{a|l} \sim x^{-2-l}, \quad a=0,1,2.
\end{equation}
Since $R^{(\mathrm{I})}_{a|l}$ diverges for $l > 0$, only $R^{(\mathrm{II})}_{a|l}$ remains well-behaved at spatial infinity, ensuring an appropriate physical decay.
The asymptotic expressions of the solutions near-horizon regime are:
\begin{subequations}
\begin{eqnarray}&&
R^{(\mathrm{I})}_{0|l}, R^{(\mathrm{I})}_{1|l} \sim \text{const.},
\quad R^{(\mathrm{I})}_{2|l} \sim (x-1);
\\&&
R_{0|l}^{(\mathrm{II})} \sim (x - 1)^{-1},
\quad R_{1|l}^{(\mathrm{II})}, \ln(x-1),
\quad R_{2|l}^{(\mathrm{II})} \sim \text{constant}.
\end{eqnarray}
\end{subequations}
Thus, only $R^{(\mathrm{I})}_{a|l}$ solutions are regular near the event horizon. For $l = 0$, only $R_{1|0}$ exists,
and among its solutions, only $R_{1|0}^{(\mathrm{II})}$ remains well-behaved in both the far-field limit ($x \to \infty$) and the near-horizon limit ($x \to 1$).
In short, to have physically acceptable solutions, we have to make the general solutions (\ref{GS_of_0R})
to have proper reasonable asymptotic behaviors both at infinity and near the horizon, namely,
$R^{(\mathrm{II})}_{a|l}$ and $R^{(\mathrm{I})}_{a|l}$ are chosen respectively.

Next, we consider the non-homogeneous case of Eq.~(\ref{LVeq_of_radial_part0}) in the presence of source terms.
We assume that the source is localized within the finite region $r_1 \leq r \leq r_2$, where $2M < r_1 < r_2 < \infty$. In this range, we may let $r_1$ sufficiently larger than the Schwarzschild radius of the compact object,
say, $r_1=10\,r_S=20M$, and $r_2$ far from infinity, \ie, $r_2<<\infty$.
This ensures that essential non-linear or curvature effects do not become dominate. More precisely, since
our analysis is primarily based on the test particle assumption, any back-reaction of electromagnetic perturbation on
the background spacetime metric, as well as potential instability issues, are beyond the scope of this work. While
these topics are indeed very interesting and important, they are significantly more complex than the current study
and are worth exploring in future researches.

Based on the previously analyzed asymptotic behaviors, we utilize the fundamental solution set $\{ R_{a|l}^{(\mathrm{I})}, R_{a|l}^{(\mathrm{II})} \}$ for $a=0,1,2$ to construct the general solution of Eq.~(\ref{LVeq_of_radial_part0}). The radial solutions in different regions are given as follows.

For $l \neq 0$, the solution takes different forms depending on the radial range:
\begin{itemize}
    \item In the region $2M < r < r_1$, where the solution is near the source but outside the event horizon, the general form is
    \begin{equation}
        R_{a|l m} = u_{l m} R_{a|l}^{(\mathrm{I})}, \quad a=0,1,2,
    \end{equation}
    where the coefficients $u_{l m}$ correspond to $a_{l m}, c_{l m}, e_{l m}$, respectively.

    \item In the region $r > r_2$, far from the source, the solution takes the form
    \begin{equation}
        R_{a|l m} = v_{l m} R_{a|l}^{(\mathrm{II})}, \quad a=0,1,2,
    \end{equation}
    where the coefficients $v_{l m}$ correspond to $b_{l m}, d_{l m}, f_{l m}$, respectively.
\end{itemize}
This piecewise formulation ensures that the solutions satisfy the appropriate boundary conditions at both spatial infinity and the event horizon while maintaining mathematical consistency across the defined radial domains.

There exists a special case for $l = 0$, where the radial function takes the form
\begin{equation}
R_{1|00} = E_{a}\, R_{1|0}^{(\mathrm{I})}, \quad \text{for} \quad 2M<r<r_1,
\end{equation}
and
\begin{equation}
R_{1|00} = E_{b}\, R_{1|0}^{(\mathrm{II})}, \quad \text{for} \quad r>r_{2}.
\end{equation}
Here, $E_{a}$ and $E_{b}$ are constants that, along with the previously introduced coefficients $ u_{l m}$ and $ v_{l m}$, will be determined using the method outlined below.

For the case of given sources in Eq.~(\ref{LVeq_of_radial_part0}), we apply the method of variation of constants \cite{ODEQ} (see also Appendix C). The corresponding particular solutions are obtained as follows:
\begin{equation}
 R^{(0)}_{a|l m}(x) = R_{a|l}^{(\mathrm{I})}(x) \int \frac{ J_{a|l m}(\xi) R_{a|l}^{(\mathrm{II})}(\xi)}{\xi(\xi-1) W\left( R_{a|l}^{(\mathrm{I})}, R_{a|l}^{(\mathrm{II})}, \xi\right)} \mathrm{d} \xi  
 -R_{a|l}^{(\mathrm{II})}(x) \int \frac{J_{a|l m}(\xi)R_{a|lm}^{(\mathrm{I})}(\xi)}{\xi(\xi-1) W\left(R_{a|lm}^{(\mathrm{I})},R_{a|lm}^{(\mathrm{II})}, \xi\right)} \mathrm{d} \xi,
\label{solution_of_0R}
\end{equation}
where $a=0,1,2$. The function $W\left( R_{a|l}^{(\mathrm{I})}, R_{a|l}^{(\mathrm{II})}, \xi\right)$ represents the Wronskian determinant of the two fundamental solutions $R_{a|l}^{(\mathrm{I})}$ and $R_{a|l}^{(\mathrm{II})}$ evaluated at $\xi$. Comparing Eq.~(\ref{solution_of_0R}) with Eq.~(\ref{GS_of_0R}), we obtain the following integral expressions for the expansion coefficients:
\begin{subequations}\label{abcdef}
\begin{eqnarray}
&& a^{(0)}_{l m}=\int_{x_1-\varepsilon}^{x_2+\varepsilon} \frac{ J_{0|l m}(x) R_{0|l}^{(\mathrm{II})}(x)}{x(x-1) W\left( R_{0|l}^{(\mathrm{I})}, R_{0|l}^{(\mathrm{II})}, x\right)} \mathrm{d} x ,\quad
b^{(0)}_{l m}=-\int_{x_1-\varepsilon}^{x_2+\varepsilon}  \frac{ J_{0|l m}(x) R_{0|l}^{(\mathrm{I})}(x)}{x(x-1) W\left( R_{0|l}^{(\mathrm{I})}, R_{0|l}^{(\mathrm{II})}, x\right)} \mathrm{d} x , \\ \nonumber \\
&& c^{(0)}_{l m}=\int_{x_1-\varepsilon}^{x_2+\varepsilon} \frac{ J_{1|l m}(x) R_{1|l}^{(\mathrm{II})}(x)}{x(x-1) W\left( R_{1|l}^{(\mathrm{I})}, R_{1|l}^{(\mathrm{II})}, x\right)} \mathrm{d} x ,\quad
d^{(0)}_{l m}=- \int_{x_1-\varepsilon}^{x_2+\varepsilon} \frac{ J_{1|l m}(x) R_{1|l}^{(\mathrm{I})}(x)}{x(x-1) W\left( R_{1|l}^{(\mathrm{I})}, R_{1|l}^{(\mathrm{II})}, x\right)} \mathrm{d} x  ,\\ \nonumber \\
&& e^{(0)}_{l m}=\int_{x_1-\varepsilon}^{x_2+\varepsilon} \frac{ J_{2|l m}(x) R_{2|l}^{(\mathrm{II})}(x)}{x(x-1) W\left( R_{2|l}^{(\mathrm{I})}, R_{2|l}^{(\mathrm{II})}, x\right)} \mathrm{d} x ,\quad
f^{(0)}_{l m}=- \int_{x_1-\varepsilon}^{x_2+\varepsilon} \frac{ J_{2|l m}(x) R_{2|l}^{(\mathrm{I})}(x)}{x(x-1) W\left( R_{2|l}^{(\mathrm{I})}, R_{2|l}^{(\mathrm{II})}, x\right)} \mathrm{d} x ,
\end{eqnarray}
\end{subequations}
where $\varepsilon$ is an infinitesimal positive constant, and $x_a\equiv r_a/(2M)$ with $a=1,2$ specify the external source region in radial direction.
After performing explicit calculations, we find that the Wronskians are approximated as follows:
\begin{subequations}\label{Wrongs1-3}
\begin{eqnarray}
&& W\left( R_{0|l}^{(\mathrm{I})}, R_{0|l}^{(\mathrm{II})}, x\right) \approx \frac{(2l+1)!}{l!(l+1)!}x^{-2}(x-1)^{-2}, \\ \nonumber \\
&& W\left( R_{1|l}^{(\mathrm{I})}, R_{1|l}^{(\mathrm{II})}, x\right) \approx \frac{2(2 l+1)!}{[(l+1)!]^{2}} x^{-3}(x-1)^{-1}, \\ \nonumber \\
&& W\left( R_{2|l}^{(\mathrm{I})}, R_{2|l}^{(\mathrm{II})}, x\right) \approx -\frac{(2l+1)!}{l!(l+1)!}x^{-4}.
\end{eqnarray}
\end{subequations}
The detailed derivation of these expressions is provided in Appendix B. Substituting Eqs. (\ref{Wrongs1-3}) into
Eqs. (\ref{abcdef}), we obtain the final expressions:
\begin{subequations}
\begin{eqnarray}
&& a^{(0)}_{l m}=\frac{l!(l+1)!}{(2 l+1)!} \int_{x_1-\varepsilon}^{x_2+\varepsilon} x(x-1)\  J_{0|l m} R_{0|l}^{(\mathrm{II})}(x) \mathrm{d} x, \\
&& b^{(0)}_{l m}=-\frac{l!(l+1)!}{(2 l+1)!} \int_{x_1-\varepsilon}^{x_2+\varepsilon} x(x-1)\  J_{0|l m} R_{0|l}^{(\mathrm{I})}(x) \mathrm{d} x, \\ \nonumber \\
&& c^{(0)}_{l m}=\frac{[(l+1)!]^{2}}{2(2 l+1)!}  \int_{x_1-\varepsilon}^{x_2+\varepsilon} x^2\  J_{1|l m} R_{1|l}^{(\mathrm{II})}(x) \mathrm{d} x, \\
&& d^{(0)}_{l m}=-\frac{[(l+1)!]^{2}}{2(2 l+1)!}  \int_{x_1-\varepsilon}^{x_2+\varepsilon} x^2\  J_{1|l m} R_{1|l}^{(\mathrm{I})}(x) \mathrm{d} x, \\ \nonumber \\
&& e^{(0)}_{l m}=-\frac{l!(l+1)!}{(2 l+1)!}  \int_{x_1-\varepsilon}^{x_2+\varepsilon} \frac{x^3}{(x-1)}\  J_{2|l m} R_{2|l}^{(\mathrm{II})}(x) \mathrm{d} x.\\ \nonumber \\
&&f^{(0)}_{l m}=\frac{l!(l+1)!}{(2 l+1)!}  \int_{x_1-\varepsilon}^{x_2+\varepsilon} \frac{x^3}{(x-1)}\  J_{2|l m} R_{2|l}^{(\mathrm{I})}(x) \mathrm{d} x .
\end{eqnarray}
\end{subequations}
For the coefficients $E_{a}$ and $E_{b}$, we refer to Eq.~(\ref{spin-s LV Maxwell}),
from which we obtain $E_{a} = 0$ and $E_{b} = \sqrt{\pi}\,e$,
where $e$ is the elementary charge obtained from integration of the charge aspect $\Phi_1^0$
of the Coulomb mode $\Phi_1$. Next, we consider the first-order correction, {\it i.e.}, Eq.~(\ref{NPLV-Maxw-1st}). Substituting the right hand side of Eq.~(\ref{01formal solution}) into Eq.~(\ref{NPLV-Maxw-1st}) and utilizing the properties of spin-weighted spherical harmonics (\ref{spheri-harmon-identy}), we obtain:
\begin{eqnarray}
&&
\hspace{-8mm}\sum_{lm}[(D-3 \rho)(\Delta+\mu-2 \gamma)-\delta(\bar{\delta}-2 \alpha)]\   R_{0|l m}^{(1)} \ {}_{1}Y_{l m} \nonumber\\&&
=-i \sum_{l m}\left[-  \frac{1}{\sqrt{2} r}[l(l+1)]^{1 / 2} k^2 \left( R^{(0)}_{1|l m} \ {}_1Y_{l m}-\bar{R}_{1|l m}^{(0)}(-1)^{m}\  {}_1 Y_{l(-m)}\right)\right.\nonumber\\&&
\left.
+(D-3 \rho) k^2 \  \bar{R}_{2|lm}^{(0)}(-1)^{-1+m}\  {}_{1}Y_{l(-m)}+(D-3 \rho) k^1\   R_{0|l m}^{(0)}\  {}_{1} Y_{l m}\right]\\&&
\hspace{-8mm}\sum_{lm}[(D-2 \rho)(\Delta  +2 \mu)-(\delta+2 \beta) \bar{\delta}] \  R_{1|l m}^{(1)} \  Y_{0|l m} \nonumber\\&&
=-i \sum_{l m} \frac{1}{\sqrt{2} r}[(l+1) l]^{1 / 2} \left( k^2\   R_{2|l m}^{(0)}\  {}_0 Y_{l m}  + k^1\   \bar{R}_{0|l m}^{(0)}(-1)^{1+m}\  {}_0 Y_{l(-m)} \right )
\nonumber\\&&
 +(D-2 \rho) k^1\left( R_{1|l m}^{(0)}\ {}_{0} Y_{l m}-(-1)^m\   \bar{R}_{1|l m}^{(0)}\  {}_0 Y_{l(-m)}\right)\\&&
\hspace{-8mm}
\sum_{lm}{[(\Delta+3 \mu)(D-\rho)-\bar{\delta}(\delta+2 \beta)]\  R_{2|l m}^{(1)}\ {}_{-1} Y_{l m} } \nonumber\\&&
=-i \sum_{l m}(\Delta+3 \mu) k^2 \  R_{2|l m}^{(0)}\ {}_{-1} Y_{l m}+(\Delta+3 \mu) k^1 \  \bar{R}_{0|l m}^{(0)} (-1)^{1+m}\ {}_{-1} Y_{l(-m)} \nonumber\\&&
 +\frac{1}{\sqrt{2} r}[l(l+1)]^{1 / 2} k^1 \left( R_{1|l m}^{(0)}\ {}_{-1}Y_{l m}-(-1)^{m}\ \bar{R}_{1|l m}^{(0)}\ { }_{-1} Y_{l(-m)}\right).
\end{eqnarray}
Using the orthogonality of spin-weighted harmonics (\ref{orthogonality condition}), we obtain the following equations for the radial components:
\begin{subequations}\label{LVeq_of_radial_part1}
\begin{eqnarray}&&
[(D-3 \rho)(\Delta+\mu-2 \gamma)-\delta(\bar{\delta}-2 \alpha)]\   R_{0|l m}^{(1)} = J^\mathrm{LV}_{0|lm},\\&&
[(D-2 \rho)(\Delta+2 \mu)-(\delta+2 \beta) \bar{\delta}] \  R_{1|l m}^{(1)}  = J^\mathrm{LV}_{1|lm},\\&&
[(\Delta+3 \mu)(D-\rho)-\bar{\delta}(\delta+2 \beta)]\  R_{2|l m}^{(1)}  = J^\mathrm{LV}_{2|lm},
\end{eqnarray}
\end{subequations}
where we define a set of effective source terms $ J_{a|l m}^\mathrm{LV}$ ($a=0,1,2$) induced by LV as following:
{\small
\begin{subequations}\label{LVcurrents}
\begin{eqnarray}&&
\hspace{-8mm}
i\,J^\mathrm{LV}_{0|lm}\equiv\left[-\frac{1}{\sqrt{2} r} k^2 [l ( l+1)]^{1 / 2}\left(R_{1|l m}^{(0)}-\bar{R}_{1|l(-m)}^{(0)}(-1)^{0-m}\right)
+(D-3 \rho)  k^2 \  \bar{R}_{2|l(-m)}^{(0)} (-1)^{-1-m}+(D-3 \rho) k^1 \  R_{0|l m}^{(0)} \right],
\\&&
\hspace{-8mm}
i\,J^\mathrm{LV}_{1|lm}\equiv\left[-\frac{1}{\sqrt{2} r}[(l+1) l]^{1 / 2}  k^2\  R_{2|l m}^{(0)}-\frac{1}{\sqrt{2} r}[(l+1) l]^{1 / 2} k^1\  \bar{R}_{0|l(-m)}^{(0)} (-1)^{1-m}
+(D-2 \rho) k^1\left( R_{1|l m}^{(0)} -(-1)^m \  \bar{R}_{1|l(-m)}^{(0)}\right)\right],
\\&&
\hspace{-8mm}
i\,J^\mathrm{LV}_{2|lm}\equiv
\left[(\Delta+3 \mu) k^2\ R_{2|l m}^{(0)}+(\Delta+3 \mu) k^1\  \bar{R}_{0|l(-m)}^{(0)} (-1)^{1-m}
+\frac{1}{\sqrt{2}r}[l(l+1)]^{1 / 2} k^1 \left( R_{1|l m}^{(0)}- \bar{R}_{1|l(-m)}^{(0)}(-1)^{0-m}\right)\right].
\end{eqnarray}
\end{subequations}
From Eqs.~(\ref{solution_of_radial_partb})-(\ref{solution_of_radial_partf}), we observe that $ R_{a|l}^{(\mathrm{II})}$ are functions of $x^{-1}$. As $x \rightarrow \infty$, we have $x^{-1} \rightarrow 0$, implying that $ R_{a|l}^{(\mathrm{II})} \rightarrow (-x)^{-2-l}$ for $a=0,1,2$. Consequently, we approximate:
\begin{eqnarray}
&& R^{(0)}_{0|l m} \approx b^{(0)}_{lm}(-x)^{-2-l}, \quad
R^{(0)}_{1|l m} \approx d^{(0)}_{lm}(-x)^{-2-l}, \quad
R^{(0)}_{2|l m} \approx f^{(0)}_{lm}(-x)^{-2-l}.
\end{eqnarray}
Substituting these approximations and the spin coefficients into Eq. (\ref{LVcurrents}) and setting $r=2Mx$,
we derive the governing radial equations (\ref{LVeq_of_radial_part1}) up to the lowest order approximations
of $R_{a|l m}^{(0)}$ and their complex conjugate $\bar{R}_{a|l m}^{(0)}$, where $a=0,1,2$.

Similar to the zeroth-order case, we express the general solution as:
\begin{equation}
\begin{aligned}
 R^{(1)}_{0|l m} & =a^{(1)}_{l m}\  R_{0|l}^{(\mathrm{I})}+b^{(1)}_{l m}\  R_{0|l}^{(\mathrm{II})}, \\
 R^{(1)}_{1|l m} & =c^{(1)}_{l m}\  R_{1|l}^{(\mathrm{I})}+d^{(1)}_{l m}\  R_{1|l}^{(\mathrm{II})},\\
 R^{(1)}_{2|l m} & =e^{(1)}_{l m}\  R_{2|l}^{(\mathrm{I})}+f^{(1)}_{l m}\  R_{2|l}^{(\mathrm{II})}.
\end{aligned}\label{GS_of_1R}
\end{equation}
For the given LV sources $ J_{a|l m}^\mathrm{LV}$ ($a=0,1,2$), following the procedure used in Eq.~(\ref{solution_of_0R}), we obtain the particular solution:
\begin{equation}\label{solution_of_1R}
\begin{aligned}
 R^{(1)}_{a|l m}(x)  = R_{a|l}^{(\mathrm{I})}(x) \int \frac{ J^\mathrm{LV}_{a|l m}(\xi) R_{a|l}^{(\mathrm{II})}(\xi)}{\xi(\xi-1) W\left(R_{a|l}^{(\mathrm{I})}, R_{a|l}^{(\mathrm{II})}, \xi\right)} \mathrm{d} \xi 
 -  R_{a|l}^{(\mathrm{II})}(x) \int \frac{ J^\mathrm{LV}_{a|l m}(\xi) R_{a|l}^{(\mathrm{I})}(\xi)}{\xi(\xi-1) W\left( R_{a|l}^{(\mathrm{I})}, R_{a|l}^{(\mathrm{II})}, \xi \right)} \mathrm{d} \xi.
\end{aligned}
\end{equation}
Comparing Eq.~(\ref{solution_of_1R}) with Eq.~(\ref{GS_of_1R}), we determine the coefficients:
\begin{subequations}
\begin{eqnarray}
&&
a^{(1)}_{l m}=\frac{l!(l+1)!}{(2 l+1)!} \int_{x_1-\varepsilon}^{x_2+\varepsilon} x(x-1)\  J^\mathrm{LV}_{0|l m}(x)\ R_{0|l}^{(\mathrm{II})}(x) \mathrm{d} x, \\
&&
b^{(1)}_{l m}=-\frac{l!(l+1)!}{(2 l+1)!} \int_{x_1-\varepsilon}^{x_2+\varepsilon} x(x-1)\ J^\mathrm{LV}_{0|l m}(x)\ R_{0|l}^{(\mathrm{I})}(x) \mathrm{d} x, \\
&&
c^{(1)}_{l m}=\frac{[(l+1)!]^{2}}{2(2 l+1)!}  \int_{x_1-\varepsilon}^{x_2+\varepsilon} x^2\ J^\mathrm{LV}_{1|l m}(x)\ R_{1|l}^{(\mathrm{II})}(x) \mathrm{d} x, \\
&&
d^{(1)}_{l m}=-\frac{[(l+1)!]^{2}}{2(2 l+1)!}  \int_{x_1-\varepsilon}^{x_2+\varepsilon} x^2\ J^\mathrm{LV}_{1|l m}(x)\ R_{1|l}^{(\mathrm{I})}(x) \mathrm{d} x, \\
&&
e^{(1)}_{l m}=-\frac{l!(l+1)!}{(2 l+1)!}  \int_{x_1-\varepsilon}^{x_2+\varepsilon} \frac{x^3}{(x-1)}\  J^\mathrm{LV}_{2|l m}(x) R_{2|l}^{(\mathrm{II})}(x) \mathrm{d} x, \\
&&
f^{(1)}_{l m}=\frac{l!(l+1)!}{(2 l+1)!}  \int_{x_1-\varepsilon}^{x_2+\varepsilon} \frac{x^3}{(x-1)}\
J^\mathrm{LV}_{2|l m}(x) R_{2|l}^{(\mathrm{I})}(x) \mathrm{d} x.
\end{eqnarray}
\end{subequations}
It can be seen that if the CPT-violating (CPTV) term is treated as an effective source term,
the Lorentz-violating (LV) effect may be characterized by the effective currents $J_{a \mid l m}^\mathrm{L V}, a=0,1,2$.
In the case of point charges or other sources (see [33]), as $x \rightarrow \infty$, the two quantities $b_{l m}^{(1)}$ and $f_{l m}^{(1)}$,
corresponding to the spin-weight $-1$ and spin-weight $+1$ modes, respectively, exhibit nearly identical asymptotic behavior.
The primary difference between these spin-weight $\mp 1$ modes arises from the LV-induced effective currents
$J_{0 \mid l m}^\mathrm{L V}$ and $J_{2 \mid l m}^\mathrm{L V}$,
which is consistent with expectations.

The analytical solutions obtained in this section reveal that CPT-odd corrections induce significant mixing between different spin-weight modes of the electromagnetic field in the farfield regime
($r \gg 2 M$), even though the electromagnetic field produced by static currents decays rapidly in this region.
In the classical Lorentz-invariant (LI) case, the zeroth-order solutions exhibit radial dependencies governed by the expansion of hypergeometric functions $R_{a \mid l}^{(0)}(x)$,
while the angular components are precisely described by spin-weighted spherical harmonics ${ }_s Y_{l m}$.
The degeneracy in the behavior of different spin-weight modes is a direct consequence of the underlying spacetime symmetry.
However, the introduction of the LV term breaks this symmetry,
manifesting as perturbations to the original solutions through the equivalent source terms $J_{a \mid l m}^\mathrm{L V}, a=0,1,2$.
Specifically, the radial behavior $R_{a \mid l m}^{(1)}$ still follows the power-law decay $x^{-l-2}$ (where $x=r / 2 M$ ),
reflecting the suppression of electromagnetic multipole radiation.
Notably, in the lowest-order case $(l=0), R_{a \mid l m}^{(1)} \propto x^{-2}$, leading to $\left|R_{a \mid l m}^{(1)}\right|^2 \sim r^{-4}$,
which confirms the absence of net energy-momentum flux in the far-field region for electromagnetic fields generated by external static currents \cite{emt}.
Furthermore, the linear relation $R_{a \mid l m}^{(1)} \propto k^a \cdot R_{a \mid l m}^{(0)}$ (for $a=1,2$ ) in the far-field solutions indicates that LV effects could be extremely small,
yet their cumulative impact might become significant in high-energy astrophysical environments, such as active galactic nucleus (AGN) jets.

Moreover, it is interesting to note that in time-dependent scenarios where radiation is present,
the two helicity $\pm$ modes correspond to different polarization states,
making helicity-dependent effects particularly pronounced for higher multipole moments ($l \geq 1$).
This suggests that LV effects may be more significant in radiation from higher multipole moments and could be constrained through cumulative effects in long-baseline photon propagation,
such as polarization angle evolution in gamma-ray bursts (GRBs).
Furthermore, the study of far-field behaviors in radiative cases reveals distinct deviations from LI electrodynamics \cite{CPTVAsymp}.
For instance, logarithmic correction terms may arise due to the absence of an additional derivative in the CPT-odd $k_{A F}$ term compared to the LI Maxwell theory \cite{Brett2020, CPTVAsymp}. Additionally, a form of energy flux cancellation between lower and higher frequency modes may occur to ensure the absence of net radiation for a charged particle moving at constant velocity \cite{Brett2015}.
For time-dependent situations, a nonzero net energy flux is expected, just as for conventional electrodynamics.
a nontrivial example is provided by dipole radiation \cite{vacuoCerenRad}, though its polarization structure is non-perturbative in terms of the CPTV coefficients.}
In fact, a closer examination of Eqs.(\ref{spin-s LV Maxwell}) reveals that the presence of the imaginary part of $\Phi_1$ in the first equation obstructs the separability of
time and radial variables, $u$ and $r$ (for further details, see Eqs. (54c) in Ref. \cite{CPTVAsymp}).
This issue may stem from the rather strong assumption of a constant radial component $k^r$.
A more physically plausible assumption --- namely, that $k^r\propto\mathcal{O}(1/r)$ --- could resolve this difficulty.
Under such a decay, the standard separability of $u$ and $r$ is recovered at sufficiently large radii, such as at null infinity.
Then the compatibility of asymptotic flatness with test particle assumption is ensured.

However, please note that current constraints on dimension-3 CPTV coefficient $k_\mathrm{AF}$ are
extremely stringent and have already attained $|k_\mathrm{AF}|\le 10^{-44}$ GeV.
For a more detailed summary of the constraints, please see tables S3 and D16 in Ref. \cite{table}.
It is also important to distinguish between Lorentz violating constraints in curved spacetime from those in flat (or conformal flat) spacetime.
The former may be subject to screen effect due to the minuscule nature of gravitational couplings \cite{Prospects2008}.
While the existing constraints are so tight that further improvements via astrophysical observations may be impractical,
our work still offers semi-analytical solutions to the CPTV Maxwell field equations in curved spacetime, which may be valuable from a theoretical standpoint.

\section{ Summary}
In this work, we investigate analytical solutions of the CPT-odd Maxwell equations in Schwarzschild spacetime using the Newman-Penrose (NP) formalism.
By employing a perturbative approach, we treat the Lorentz-invariant (LI) Maxwell equations as the zeroth-order approximation and incorporate the CPT-violating (CPTV) coefficients
$\left(k_{A F}\right)^\mu$ as first-order corrections.
The electromagnetic field tensor is decomposed into three complex NP scalars $\Phi_0, \Phi_1, \Phi_2$, whose radial dependence is governed by hypergeometric functions,
while the angular components are described by spin-weighted spherical harmonics.
Specifically, the zeroth-order solutions preserve the spherical symmetry of the Schwarzschild metric, with multipole moments determined by hypergeometric radial functions.
The introduction of CPTV coefficients $\left(k_{A F}\right)^\mu$ induces anisotropic corrections by coupling different angular modes $(l, m)$,
even assuming the CPTV coefficients are spherical symmetrically distributed.
In the radiation case, this may also alter the polarization structure.
If the radial behavior of radiation follows a similar scaling $\left(R_{a \mid l m}^{(1)} \sim x^{-l-1}\right)$ with $x=r / 2 M$,
it may indicate the suppression of higher multipole radiation at large distances.

Although this is a very preliminary study of CPT-violating electrodynamics in curved spacetime,
it represents a first attempt within the Newman-Penrose formalism.
The interplay between geometric and Lorentz-violating (LV) effects may not only deepen our understanding of black hole electrodynamics
but also open new avenues for exploring physics beyond classical relativity.
From a theoretical perspective, the results presented here provide a foundation for further investigations into CPT-violating and LV effects in curved spacetime.
Future research could explore more complex spacetime backgrounds, such as Kerr spacetime, where frame-dragging and ergo-region dynamics may significantly influence LV effects.
Extending this analysis to rotating black holes could reveal novel phenomena,
such as LV-modified superradiance or potential imprints on black hole shadow substructures observable by the Event Horizon Telescope.
Moreover, the framework developed here is inherently adaptable to arbitrary orders of electric and magnetic multipole expansions.
By extending the relation $R_{a \mid l m}^{(1)} \propto x^{-l-2}$ to higher $l$,
one could systematically analyze LV corrections to higher multipole moments, probing finer details of electromagnetic fields near compact objects.

\section{Acknowledgment}
We would like to appreciate the valuable discussion with Zhanfeng Mai and H. Wang want to express his gratitude to Tengfei Li for his assistance during the calculation process.
This work was sponsored by National Natural Science Foundation of China under grant No. 11605056, and Bing Sun was sponsored by National Natural Science Foundation of China under grant No. 12375046
and No 12475046, and Beijing University of Agriculture QJKC-2023032.

\section{Appendix}
\subsection{ Spin-weighted spherical harmonics}
The spin-weighted spherical harmonics could be defined in terms of the usual spherical harmonics as:
\begin{equation}
{ }_s Y_{l m}=
\begin{cases}\sqrt{\frac{(l-s)!}{(l+s)!}} \eth^s Y_{l m}, & 0 \leq s \leq l \\ \sqrt{\frac{(l+s)!}{(l-s)!}}(-1)^s \bar{\eth}^{-s} Y_{l m}, & -l \leq s \leq 0 \\ 0, & l<|s| .\end{cases}
\end{equation}
It could also be represented as:
\begin{equation}
\begin{aligned}
{ }_s Y_{l m}(\theta, \phi)=&(-1)^{l+m-s} \sqrt{\frac{(l+m)!(l-m)!(2 l+1)}{4 \pi(l+s)!(l-s)!}} \sin ^{2 l}\left(\frac{\theta}{2}\right) e^{i m \phi} \times \\
& \times \sum_{r=0}^{l-s}(-1)^r\binom{l-s}{r}\binom{l+s}{r+s-m} \cot ^{2 r+s-m}\left(\frac{\theta}{2}\right) .
\end{aligned}
\end{equation}
From the definition above, the spin-weighted spherical harmonics have some useful properties
\begin{eqnarray}\label{spheri-harmon-identy}&&
{ }_s \bar{Y}_{l m}  =(-1)^{m+s}{ }_{-s} Y_{l(-m)}, \qquad
{ }_1 Y_{l m} ={ }_{-1} Y_{l m}+2 m[l(l+1)]^{-1 / 2}(\sin \theta)^{-1} Y_{l m},\qquad
\frac{\partial}{\partial \varphi}\ { }_s Y_{l m} =\mathrm{i} m_s Y_{l m} \nonumber\\&&
\eth\ { }_s Y_{l m} =[(l-s)(l+s+1)]^{1 / 2}{ }_{s+1} Y_{l m}, \qquad~~~
\bar{\eth}\ {}_s Y_{l m} =-[(l+s)(l-s+1)]^{1 / 2}{ }_{s-1} Y_{l m}, \nonumber\\&&
\bar{\eth} \eth\ {}_s Y_{l m} =-(l-s)(l+s+1)_s Y_{l m}, \qquad~~~~~~~
\eth { }\bar{\eth}\ {}_s Y_{l m} =-(l+s)(l-s+1)_s Y_{l m},
 ,
\end{eqnarray}
The spin-weighted spherical harmonics obey the orthogonality condition
\begin{equation}
\int_0^{2 \pi} \int_{0}^\pi{ }_s \bar{Y}_{l m}(\theta, \varphi)_s Y_{l^{\prime} m^{\prime}}(\theta, \varphi) \mathrm{d} \Omega=\delta_{l l^{\prime}} \delta_{m m^{\prime}},\label{orthogonality condition}
\end{equation}
where $ \mathrm{d} \Omega=\cos\theta\,\mathrm{d}\theta\,\mathrm{d}\phi$.

\subsection{Wronskian}
The Wronskian is a determinant constructed from $n$ differentiable functions $f_1, \ldots, f_n$ along with their first $n-1$ derivatives. Its explicit form is given by:
\begin{equation}
W\left(f_1, \ldots, f_n\right)(x)=\operatorname{det}\left[\begin{array}{cccc}
f_1(x) & f_2(x) & \cdots & f_n(x) \\
f_1^{\prime}(x) & f_2^{\prime}(x) & \cdots & f_n^{\prime}(x) \\
\vdots & \vdots & \ddots & \vdots \\
f_1^{(n-1)}(x) & f_2^{(n-1)}(x) & \cdots & f_n^{(n-1)}(x)
\end{array}\right],
\end{equation}
where $f_n^{(n-1)}$ is $(n-1)^\mathrm{th}$ derivative
of $f_n$.
For the general homogeneous differential equation
\begin{equation}
	f^{(n)}+a_1(x)f^{(n-1)}+\cdots+a_n(x)f=0,\label{GHDE}
\end{equation}
there exists a useful relation known as Abel's identity, which expresses the Wronskian of a set of solutions in terms of a known Wronskian at a reference point and the coefficients of the original differential equation:
\begin{equation}
W\left(f_1, \ldots, f_n\right)(x)=W\left(f_1, \ldots, f_n\right)(x_0) \exp \left( - \int_{x_0} ^{x} a_{n-1}(\xi)\mathrm{d}\xi   \right).
\end{equation}
This identity is particularly useful in computing the Wronskian of $\Phi_0, \Phi_1, \Phi_2$. For $\Phi_0$, using Eq.~(\ref{HGeq0}), we obtain:
\begin{equation}
\begin{aligned}
W\left( R_{0|l}^{(\mathrm{I})},R_{0|l}^{(\mathrm{II})}, x\right) &= W\left(R_{0|l}^{(\mathrm{I})},R_{0|l}^{(\mathrm{II})}, x_0\right) \exp \left[-\int_{x_0}^x \frac{4 \xi-2}{\xi(\xi-1)} \mathrm{d} \xi\right] \\
&= W\left(R_{0|l}^{(\mathrm{I})},R_{0|l}^{(\mathrm{II})}, x_0\right) \frac{x_0^2\left(x_0-1\right)^2}{x^2(x-1)^2}.
\end{aligned}
\end{equation}
Taking the limit $x_0\rightarrow \infty$, and substituting $R_{0|l}^{(\mathrm{I})}$ and $R_{0|l}^{(\mathrm{II})}$ from Eq.~(\ref{solution_of_radial_partb}), we obtain:
\begin{eqnarray}&&
W\left(R_{0|l}^{(\mathrm{I})},R_{0|l}^{(\mathrm{II})}, x_0\right)\approx
\frac{(2l+1)!}{l!(l+1)!}x_0^{-4},\qquad
W\left(R_{0|l}^{(\mathrm{I})},R_{0|l}^{(\mathrm{II})}, x\right)\approx
\frac{(2l+1)!}{l!(l+1)!}\frac{1}{x^2(x-1)^2}.
\end{eqnarray}

For $\Phi_1$, using Eq.~(\ref{HGeq1}):
\begin{equation}
\begin{aligned}
W\left(R_{1|l}^{(\mathrm{I})},R_{1|l}^{(\mathrm{II})}, x\right) &= W\left(R_{1|l}^{(\mathrm{I})},R_{1|l}^{(\mathrm{II})}, x_0\right) \exp \left[-\int_{x_0}^x \frac{4 \xi-2}{\xi(\xi-1)} \mathrm{d} \xi\right] \\
&= W\left(R_{1|l}^{(\mathrm{I})},R_{1|l}^{(\mathrm{II})}, x_0\right) \frac{x_0^3\left(x_0-1\right)}{x^3(x-1)}.
\end{aligned}
\end{equation}
Taking the limit $x_0\rightarrow \infty$, and substituting $R_{1|l}^{(\mathrm{I})}$ and $R_{1|l}^{(\mathrm{II})}$ from Eq.~(\ref{solution_of_radial_partd}), we obtain:
\begin{eqnarray}&&
W\left(R_{1|l}^{(\mathrm{I})},R_{1|l}^{(\mathrm{II})}, x_0\right)
\approx \frac{2(2l+1)!}{[(l+1)!]^2}x_0^{-4},\qquad
W\left(R_{1|l}^{(\mathrm{I})},R_{1|l}^{(\mathrm{II})}, x\right)
\approx \frac{2(2l+1)!}{[(l+1)!]^2}\frac{1}{x^3(x-1)}.
\end{eqnarray}

For $\Phi_2$, using Eq.~(\ref{HGeq2}):
\begin{equation}
\begin{aligned}
W\left(R_{2|l}^{(\mathrm{I})},R_{2|l}^{(\mathrm{II})}, x\right) &= W\left(R_{2|l}^{(\mathrm{I})},R_{2|l}^{(\mathrm{II})}, x_0\right) \exp \left[-\int_{x_0}^x \frac{4 \xi-4}{\xi(\xi-1)} \mathrm{d} \xi\right] \\
&= W\left(R_{2|l}^{(\mathrm{I})},R_{2|l}^{(\mathrm{II})}, x_0\right) \frac{x_0^4}{x^4}.
\end{aligned}
\end{equation}
Taking the limit $x_0\rightarrow \infty$, and substituting $R_{2|l}^{(\mathrm{I})}$ and $R_{2|l}^{(\mathrm{II})}$ from Equation (\ref{solution_of_radial_partf}), we obtain:
\begin{eqnarray}&&
W\left(R_{2|l}^{(\mathrm{I})},R_{2|l}^{(\mathrm{II})}, x_0\right)\approx-\frac{(2l+1)!}{l!(l+1)!}x_0^{-4},\qquad
W\left(R_{2|l}^{(\mathrm{I})},R_{2|l}^{(\mathrm{II})}, x\right)\approx-\frac{(2l+1)!}{l!(l+1)!}\frac{1}{x^4}.
\end{eqnarray}

\subsection{The solution of non-homogeneous equation }
In this subsection, we introduce an approach~\cite{ODEQ} for solving non-homogeneous differential equations. Consider a general non-homogeneous ordinary differential equation of the form:
\begin{equation}
L(y) = y^{(n)} + a_1(x) y^{(n-1)} + \cdots + a_n(x) y = b(x), \label{non-homo ODE}
\end{equation}
where $L(y)$ represents a linear differential operator. The general solution of Eq.~(\ref{non-homo ODE}) can be written as:
\begin{equation}
y = y_p + c_1 y_1 + \cdots + c_n y_n,
\end{equation}
where $y_p$ is a particular solution of (\ref{non-homo ODE}). When $b(x) = 0$, the equation reduces to a homogeneous form, whose general solution is given by $c_1 y_1 + \cdots + c_n y_n$, where $c_1, \cdots, c_n$ are constants.

To construct a particular solution $y_p$ with a structure similar to the homogeneous solution, we assume the ansatz:
\begin{equation}
y_p = u_1 y_1 + \cdots + u_n y_n,
\end{equation}
where $u_1, \cdots, u_n$ are functions rather than constants. These functions satisfy the following system of equations:
\begin{equation}
\begin{aligned}
&\begin{aligned}
u_1^{\prime} y_1 + \cdots + u_n^{\prime} y_n & = 0, \\
u_1^{\prime} y_1^{\prime} + \cdots + u_n^{\prime} y_n^{\prime} & = 0,
\end{aligned}\\
&\begin{aligned}
u_1^{\prime} y_1^{(n-2)} + \cdots + u_n^{\prime} y_n^{(n-2)} &= 0, \\
u_1^{\prime} y_1^{(n-1)} + \cdots + u_n^{\prime} y_n^{(n-1)} &= b(x).
\end{aligned} \label{eq sets}
\end{aligned}
\end{equation}
This system of equations can be solved using Cramer's rule, yielding:
\begin{equation}
u_k(x) = \int_{x_0}^{x} \frac{W_k(t)}{W(y_1,\cdots,y_n)(t)} dt, \quad (k=1,\cdots,n),
\end{equation}
where $W(y_1,\cdots,y_n)(t)$ denotes the Wronskian determinant of the fundamental solutions. Consequently, the particular solution $y_p(x)$ can be expressed as:
\begin{equation}
y_p(x) = \sum_{k=1}^n y_k(x) \int_{x_0}^x \frac{W_k(t) b(t)}{W\left(y_1, \cdots, y_n\right)(t)} d t.
\end{equation}

\def\cmag{\color{magenta}}

\end{document}